\documentclass[letterpaper]{article} 

\usepackage{aaai23}  
\usepackage{times}  
\usepackage{helvet}  
\usepackage{courier}  
\usepackage[hyphens]{url}  
\usepackage{graphicx} 
\usepackage{natbib}  
\usepackage{caption} 
\DeclareCaptionStyle{ruled}{labelfont=normalfont,labelsep=colon,strut=off} 
\usepackage{algorithm}
\usepackage{algorithmic}

\usepackage{newfloat}
\usepackage{listings}
\floatstyle{ruled}
\newfloat{listing}{tb}{lst}{}
\floatname{listing}{Listing}
\usepackage{xcolor}

\title{Auditing Elon Musk's Impact on Hate Speech and Bots}
\author {
    Daniel Hickey,\textsuperscript{\rm 1}
    Matheus Schmitz, \textsuperscript{\rm 2}
    Daniel Fessler, \textsuperscript{\rm 3,4,5}
    Paul E. Smaldino, \textsuperscript{\rm 6,7}
    Goran Muric, \textsuperscript{\rm 2}
    Keith Burghardt \textsuperscript{\rm 2}
}

\affiliations {
    \textsuperscript{\rm 1}  {Oregon State University}\\
\textsuperscript{\rm 2} {USC Information Sciences Institute}\\
\textsuperscript{\rm 3} {University of California, Los Angeles}\\
\textsuperscript{\rm 4} {Bedari Kindness Institute, University of California, Los Angeles}\\
\textsuperscript{\rm 5} {Center for Behavior, Evolution, \& Culture, University of California, Los Angeles}\\
\textsuperscript{\rm 6} {University of California, Merced}\\
\textsuperscript{\rm 7} {Santa Fe Institute}\\
hickeyda@oregonstate.edu
}

\usepackage{bibentry}

\begin{document}

\maketitle


\begin{abstract}
On October 27th, 2022, Elon Musk purchased Twitter, becoming its new CEO and firing many top executives in the process. Musk listed fewer restrictions on content moderation and removal of spam bots among his goals for the platform. Given findings of prior research on moderation and hate speech in online communities, the promise of less strict content moderation poses the concern that hate will rise on Twitter. We examine the levels of hate speech and prevalence of bots before and after Musk's acquisition of the platform. We find that hate speech rose dramatically upon Musk purchasing Twitter and the prevalence of most types of bots increased, while the prevalence of astroturf bots decreased.
\end{abstract}

\section{Introduction}
On October 27th, 2022, Elon Musk, CEO of SpaceX and Tesla, acquired Twitter for \$44 billion USD, becoming the new CEO and firing many of its top executives \cite{osullivan2022elon}. The majority of its remaining workforce left the company in the month following Musk's purchase, through both layoffs and resignations \cite{mac2022resignations}. Musk claimed he bought the platform because of its potential as a global online platform that allows free speech \cite{sato2022buying}. Prior to the deal, Musk proposed bringing multiple changes to the platform, including making its algorithms open-source, loosening content moderation policies, and reducing the presence of spam bots on the platform \cite{sherman2022elon}. Immediately following the purchase, several reductions in content moderation occurred, including the reinstatement of Donald Trump's account \cite{duffy2022elon}, the removal of a COVID-19 misinformation policy \cite{osullivan2022twitter}, and the disbanding of the Trust and Safety Council, a board of organizations that provided guidance on content moderation \cite{obrien2022musk}. Despite these changes, on November 23rd, Musk claimed hate speech impressions dropped to a level lower than before his purchase of the platform, following an initial spike directly after the acquisition. However, he did not reveal the methodology used for hate speech detection \cite{frenkel2022hate}. Alongside policy changes on content moderation, Musk announced that efforts to combat spam bots have been enacted since the acquisition \cite{anand2022twitter}.

Given Musk's promises of less restrictive content moderation, as well as the findings of previous research that indicate lighter moderation is associated with increased hate speech on social media platforms \cite{zannettou2018gab, Chandrasekharan2017}, we hypothesize that hate speech on Twitter increased following his acquisition. We use hateful keywords combined with the Perspective API \cite{Jigsaw2017} to extract hateful tweets. We examine increases in hate speech in two ways. First, we extract timelines of a sample of users who posted hateful tweets one month before and after Musk's purchase and measure their daily rates of hate speech during the same time period. Then, we measure the overall volume of hateful tweets throughout 2022. Additionally, as Musk highlighted a reduction in bot accounts as one of his goals for the platform, we hypothesize that the prevalence of bots decreased following his acquisition, and assess whether the prevalence of bots on Twitter changed following his takeover. We determine bot scores for accounts that posted hateful tweets and accounts from the baseline sample using Botometer \cite{sayyadiharikandeh2020detection} and compare distributions of accounts' bot scores during two months before and after Musk's takeover.

Overall, we find both hate speech and bot scores increased following Musk's purchase. While no direct causal relationships between specific changes to Twitter policies or features can be inferred from our analysis, we speculate as to the likely causes of the patterns we document, and discuss implications of these results for the future of Twitter.

\section{Related Work}

\subsection{Hate Speech and Content Moderation}

Previous literature comparing social media platforms with different content moderation policies can inform how Musk's relaxation of moderation will impact Twitter. The social media platform Gab, which was created as an alternative to Twitter with fewer restrictions on speech, was found to have much higher levels of hate speech than Twitter while having lower hate speech levels than the Politically Incorrect (/pol/) board on 4Chan, a platform with notoriously light content moderation \cite{zannettou2018gab, hine2017kek}. Moderation on mainstream platforms successfully decreases hate speech or the growth of hateful communities within those platforms \cite{Chandrasekharan2017, chandrasekharan2020quarantined}. Some warn it can increase hate speech and toxicity on fringe social media platforms due to platform migration \cite{johnson2019hidden, horta2021platform}.

\subsection{Bots on Twitter}

Social bots have a significant presence on Twitter, estimated to be 9--15\% of accounts \cite{varol2017online}. Research found people struggle distinguishing between bots and humans, meaning the presence of bots on Twitter may not be obvious to its user base \cite{cresci2017paradigm}. Bots can take many forms - for example, ``astroturf'' bots artificially inflate support for political candidates, and smear their opponents \cite{ferrara2016rise}. Other types of bots include fake followers, used to build follower counts of profiles, and spammers, which generate large amounts of posts promoting certain products or viewpoints \cite{yang2022botometer, sayyadiharikandeh2020detection}. Bots pose risks to social media platforms by spreading misinformation and hate \cite{uyheng2020bots, wang2018era}. Bots have been found to play a role in discussions of many topics, including COVID-19 \cite{shi2020social}, climate change \cite{marlow2021bots}, and the stock market \cite{cresci2019cashtag}.

\section{Methods}

\subsection{Hateful tweet extraction}

To measure changes in hate speech levels following Musk's purchase of Twitter, we devise a method to identify hate speech on the platform. First, we leverage vocabulary associated with hate groups to identify a set of hateful keywords with which to pinpoint hate tweets. Given Reddit's natural group-centricity, we use known Reddit hate communities as the basis for the hateful vocabularies, presuming that the words extracted will be hateful independent of the platform on which they appear. We start with hate speech lexicons extracted from common keywords studied in Schmitz et al. \citeyear{Schmitz2022}.  We then replicate their methods for hateful keyword extraction in an additional set of 21 hate forums, known as subreddits, by collecting the top 100 keywords from each subreddit using SAGE \cite{eisenstein2011sparse}. Three human annotators then rated each keyword as 0 $=$ not hateful, 1 $=$ sometimes hateful, or 2 $=$ always hateful. Words with a cumulative score of 4 or higher were considered hate words. Words below this threshold were removed from analysis. The usage of the hateful keywords on Twitter was manually verified to ensure they had not taken on different meanings compared to their usage on Reddit, with differing words being removed. This resulted in a total of 49 keywords. The hate keywords and subreddits are found here (WARNING: Contains offensive terms): {https://github.com/dan-hickey1/musk-hate-lexicon}. Using these keywords, we collect historical tweets from the Twitter API for Academic Research. As there is a limit to how many tweets a user of the Twitter API can collect, we opt to sub-sample hateful tweets that appear on the platform during 2022 by requesting all tweets that contain terms from our hate speech lexicon created during five randomly sampled five-minute intervals each day. The intervals are different each day to ensure we did not select times of the day more or less likely to contain hate speech, though this results in a noisier estimate overall.

As the presence of hate words does not guarantee that the text is hateful, we utilize the Perspective API, a family of models used for detecting toxic content, to further filter our data \cite{Jigsaw2017}. We focus on tweets that might threaten the future usability of Twitter, so we choose to filter on Perspective's ``toxicity'' metric wherein a toxic comment is ``a rude, disrespectful, or unreasonable comment that is likely to make people leave a discussion.'' While the Perspective API was trained on New York Times data, it has been verified on a range of social media platforms, including Twitter \cite{saveski2021structure}. We consider tweets with a toxicity probability \textgreater 0.7 hateful. Many high-toxicity tweets were pornographic, and not necessarily reflective of the hate speech we seek to measure. We therefore also use Perspective's ``sexually explicit'' model to filter out this type of content. We removed tweets with a sexually explicit probability \textgreater 0.3. 
We checked the results of our analyses for varying combinations of thresholds for both metrics, as well as removing the thresholds entirely. The results are qualitatively similar.

To ensure fluctuations in hate speech are not reflective of fluctuations in overall user activity, we also sample a baseline set of tweets collected during the same time intervals. We used the keyword ``thing,'' taken from a set of the most commonly used words \cite{Leech2001}. Ideally, our keyword should be common enough to be representative of the activity on Twitter as a whole, though a set of terms that is too popular will quickly exhaust Twitter API data collection limits. As the choice of the term is somewhat arbitrary, we validated it with two other terms from the list, ``tell'' and ``any,'' finding that the weekly volumes of the terms correlated strongly with the ``thing'' keyword, with Pearson correlations of 0.84 and 0.76, respectively.

\begin{table}[h]
    \centering
    \begin{tabular}{p{60pt}llp{55pt}}
    Sample                  & Tweets & Users & Botometer Users\\ \hline
    Hateful (full year)     & 4,437            & 4,259    & 4259   \\
    Hateful (user timeline) & 143,542          & 195 &--            \\
    Baseline                & 4,921,740     & 3,092,173&2,677; 7,907 
    \end{tabular}
    \caption{Summary of data statistics}
    \label{tab:tweet_counts}
\end{table}

\subsection{Hateful user timeline collection}

To understand how individual hateful users responded to Musk's acquisition, we sampled historical tweets from the Twitter API for Academic Research, using the hate speech lexicon described in Section 3.1, at 600 randomly sampled 30-second intervals between October 1st and November 29th. Tweets were filtered using the same Perspective metrics and thresholds described in Section 3.1. We then collected all tweets from October 1st to November 29th from each author. Accounts that posted the maximum of 3,200 tweets available via the Twitter API after October 1st were removed from the sample, as we could not extract their full history. 
The number of tweets and users in each sample are listed in Table~\ref{tab:tweet_counts}. Proportions of hate speech in each users' timelines were then measured in daily bins to track trends.

\subsection{Bot detection}

To identify bots, we use Botometer, which considers over 1,000 features collected in an ensemble of random forest classifiers to predict if Twitter accounts are automated \cite{sayyadiharikandeh2020detection}. Botometer provides an overall score and predictions for different types of bots, including spambots, astroturf bots, and fake followers. We follow recommended guidelines for using Botometer, though some issues, such as the transient nature of Botometer scores, are unavoidable in the context of our current study \cite{yang2022botometer}.

We assess the Botometer scores for all accounts that posted hateful tweets. As the Botometer API has rate limits, we collected Botometer scores of a random sample of accounts before (2,677) and after (7,907) Musk's acquisition from the control sample (see Table~\ref{tab:tweet_counts} for a summary of all data collected). Using the Mann-Whitney U test, we compare the distributions of each type of bot scores of accounts that posted two months before and after Musk's purchase.

\section{Results}

\textbf{Hateful users employed more hate speech following Elon Musk's purchase of Twitter.}

\begin{figure}[h]
    \centering
    \includegraphics[width=0.85\columnwidth]{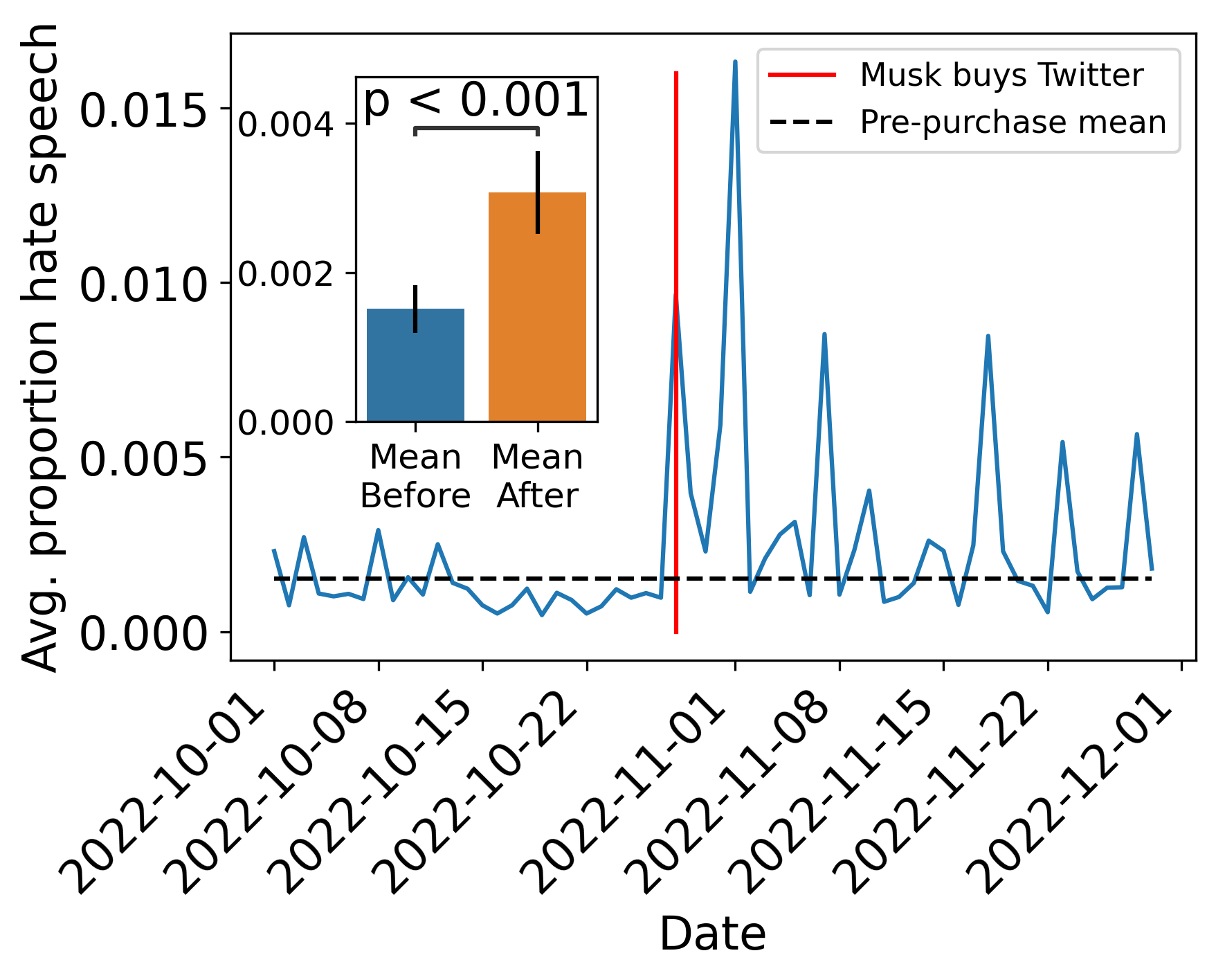}
    \caption{Hateful users increased their hate speech after Musk bought Twitter. Proportion of hate words in hateful users' tweets over time before and after the purchase. The inset plot represents the average daily hate speech before and after the takeover. Black lines in inset plot represent standard errors; p-value calculated from a Mann-Whitney U test.}
    \label{fig:user_level_hate}
\end{figure}

Among the set of users who posted hateful tweets from Oct. 1st to Nov. 29th, the average proportion of hate speech used per day increased dramatically upon Musk's purchase (Fig~\ref{fig:user_level_hate}). \emph{Hate speech levels immediately after Musk's purchase quadruple}. Qualitatively, the spike in the proportion of hate speech is robust to changes in toxicity and sexually explicit thresholds when selecting users to include in the sample. After the initial spike in hate speech, the levels of hate speech tend to still reach levels higher than the highest levels from the month before, hinting at a new baseline level of hate speech post-Musk.

\textbf{The overall presence of hateful tweets increased.}

\begin{figure}[h]
    \centering
    \includegraphics[width=0.9\columnwidth]{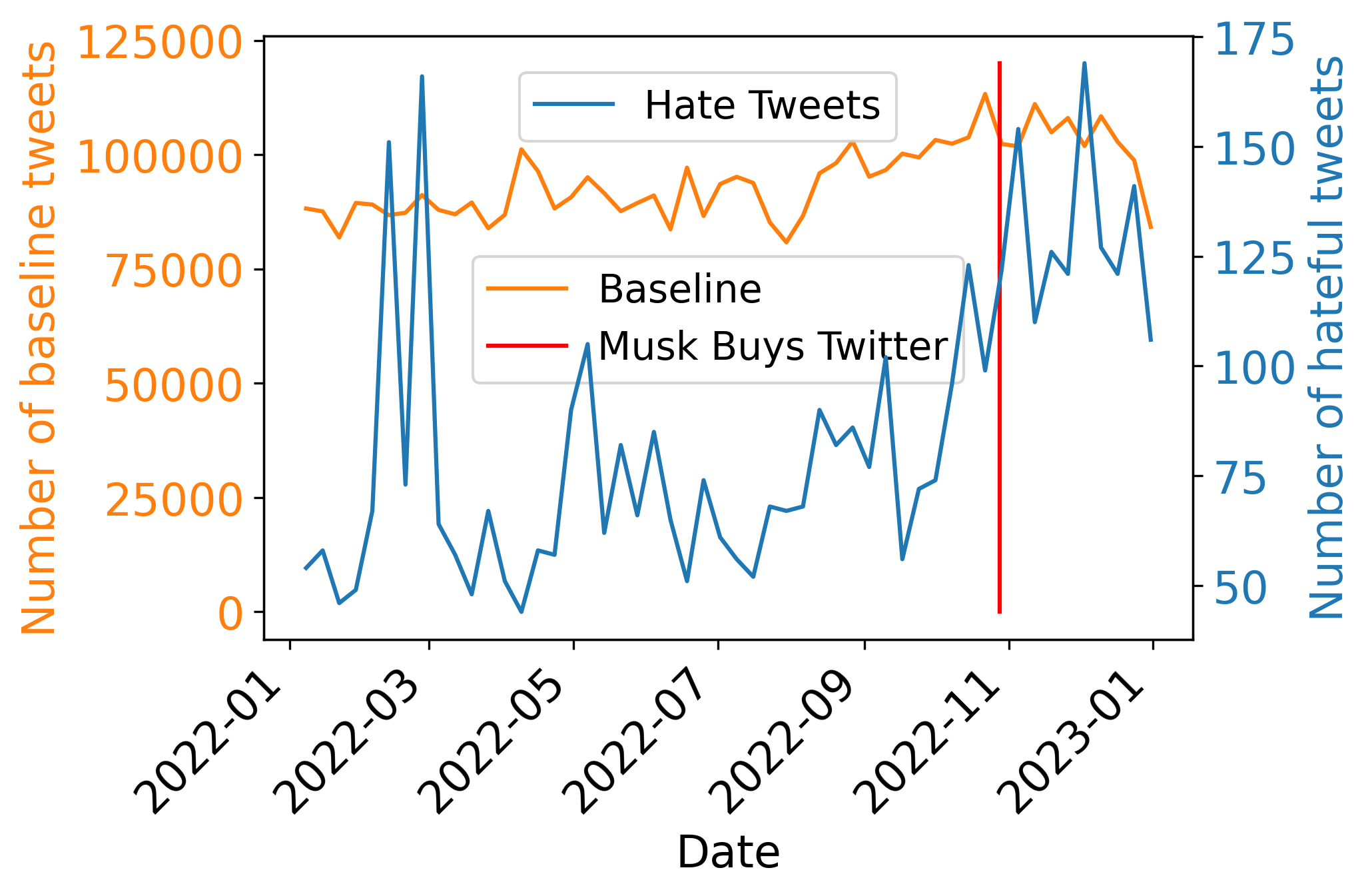}
    \caption{Overall hate speech on Twitter increased after Elon Musk bought the platform. Number of hateful tweets sampled during each week in 2022, compared to a baseline of tweets collected during the same time period. The spike in March, 2022 coincides with the Canada convoy protests.}
    \label{fig:hate_tweet_time_series}
\end{figure}

Fig.~\ref{fig:hate_tweet_time_series} displays the volume of hateful and baseline tweets sampled in 2022. While baseline tweets remain stable throughout the year (with an exception in early 2022 coinciding with the Canada convoy protests \cite{GlobalNews2022}), hateful tweets increase dramatically following Musk's takeover. This increase is robust to different combinations of thresholds for toxicity and sexually explicit content.

\textbf{The prevalence of most types of bots changed.}

\begin{figure*}[h]
    \centering
    \includegraphics[width=0.75\textwidth]{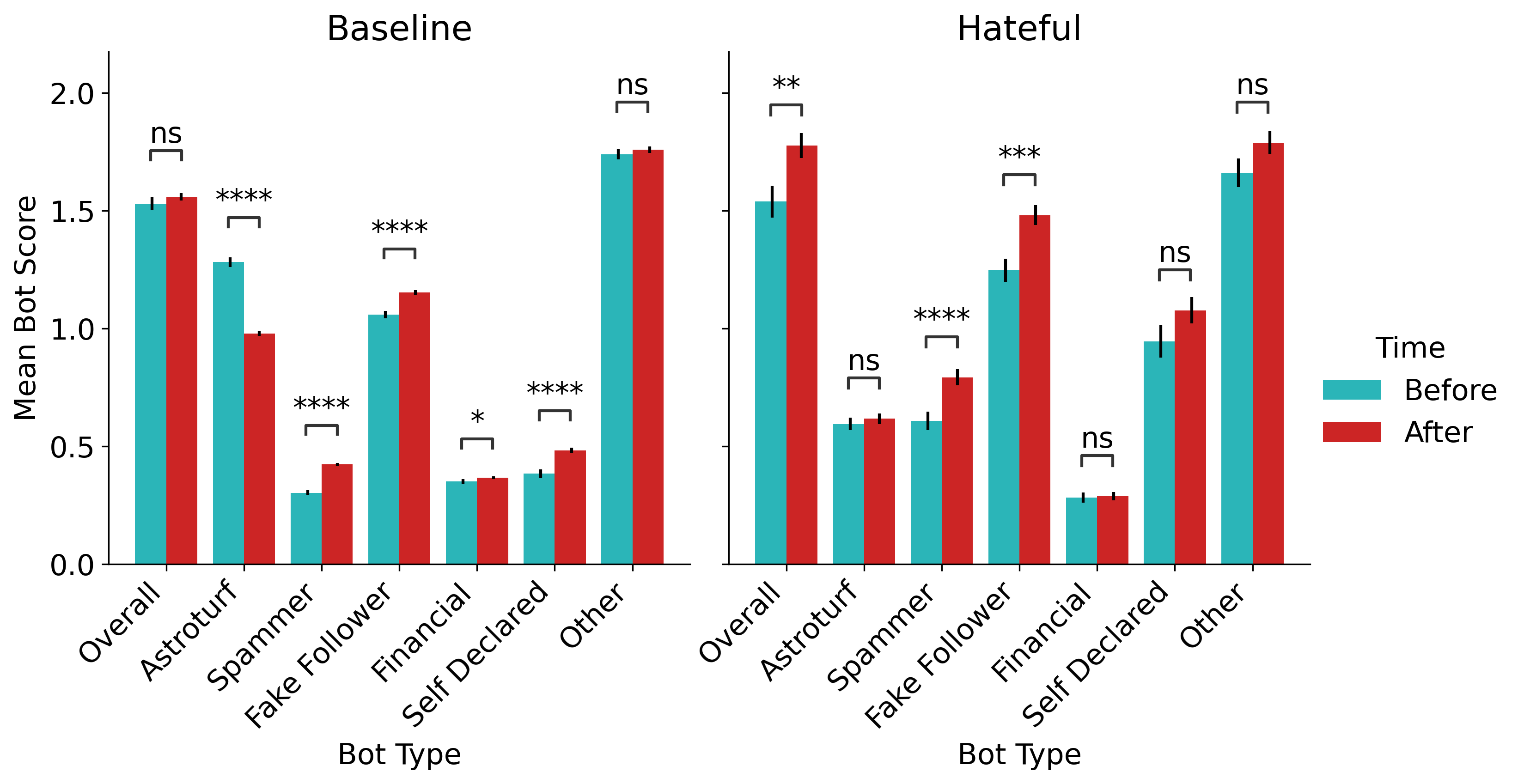}
    \caption{Mean Botometer scores before and after Musk's purchase. Black lines represent standard errors. P-values are calculated from the two-sided Mann-Whitney U test. *:$p < 0.05$, **:$p < 0.01$, ***:$p < 0.001$, ****:$p < 0.0001$}
    \label{fig:bot_bars}
\end{figure*}

Fig.~\ref{fig:bot_bars} displays relative changes in mean Botometer scores for accounts that posted in two-month periods before and after Musk's purchase. Overall, spammer, and fake follower scores increased for both hateful and control accounts, while astroturf scores decreased for the control accounts.

As new interventions to combat spam bots require time to be introduced following Musk's takeover, we checked for differences in distributions of Botometer scores in the period Oct. 27th -- Dec. 2nd compared to Dec. 3rd -- Dec. 31st. However, we found no decreases in overall bot scores, rather there were significant increases in certain categories of scores. We caution that this should not be taken as auditing the efficacy of specific bot-prevention policies.

\section{Discussion}

\textbf{Potential consequences of increased hate on Twitter.}

Both analyses we performed show large increases in hate speech following Musk's purchase, with no sign of hate speech returning to  previously typical levels. Prior research highlights the consequences of online hate speech, including increased anxiety in users \cite{saha2019prevalence} and offline victimization of targeted groups \cite{lewis2019online}. The effects of Twitter's moderation policies are thus likely far-reaching and will lead to negative consequences if left unchecked.

\textbf{Is Musk effectively reducing bot presence on Twitter?} We lack sufficient information about anti-bot actions taken on Twitter to address their individual efficacy. Musk says his goal is to reduce ``spam bots'', but it is unclear if his definition aligns with Botometer's. Musk may be particularly concerned about astroturf bots, which could explain their decreased presence. This decrease is especially interesting as it happened during the U.S. Midterm elections, presumably a period of heightened political bot activity.

A likely explanation for the increase in most types of bots is a lack of bot moderation due to the large reduction in Twitter's workforce following Musk's purchase. If workers who normally remove bot accounts resigned from Twitter or were laid off, then bot accounts would be allowed free reign.

\subsection*{Limitations and Future Directions}

We document marked changes in hate speech and bot levels on Twitter following Musk's takeover. However, several steps remain to understand the future of the platform. Below, we highlight where our current understanding of the situation is lacking and how it can be improved.

\textbf{Causal analysis and effects of specific policies.}
Our work should not be interpreted as an assessment of specific policy changes Musk instituted. Our methods cannot prove a causal relationship between Musk's takeover and hate speech or bot levels. Musk's changes to moderation and bot removal are poorly documented, so determining their effects would be difficult. However, it would be valuable to understand how specific policies have changed the platform.

\textbf{Changes in Botometer scores for individual users.} While we collect data from accounts throughout 2022, all Botometer scores were calculated after Musk's purchase of the platform. As Botometer scores are calculated based on the most recent information in a user's profile \cite{yang2022botometer}, it is possible user behaviors changed to become more bot-like following Musk's purchase, which may not be reflective of their behavior when they authored their tweets. Additionally, some accounts may have been banned or deleted after they issued the tweets collected but before we assessed Botometer scores. If an account is banned or deleted, we cannot calculate its Botometer score. Bots being banned after Musk's purchase could decrease our estimates of bots prior to Musk's purchase.

\textbf{Visibility of bots on Twitter.} While bot scores hint at the overall presence of bots on the platform, we do not know the visibility of bots on Twitter. Certain Twitter policies may have been taken to prevent users from seeing and interacting with bots without removing accounts. Further analyses on bot engagement metrics would have to be done to determine if visibility has changed following Musk's takeover.

\textbf{Musk's impact on human activity on Twitter.} As our analysis of bot activity is relative, it is possible the increase in Botometer scores is due to humans leaving the platform rather than more bots joining. This is plausible, as many users claimed they would leave the platform \cite{tiffany2022are}. Further research must be done to accurately determine if our findings are caused by people migrating from Twitter.

\section*{Broader Perspective, Ethics and Competing Interests}
All data were collected from the public Twitter API; identifiable information was removed prior to analysis, minimizing risks to Twitter users. Our work provides several potential benefits for society, including an audit of the steps ostensibly being taken to combat harm on Twitter, and a new way to detect hate speech at scale using commercial APIs as well as a curated list of hate words. Perspective API, which we use to classify hate, is run by Alphabet, a competitor to Twitter, but we believe this does not affect our results.

\section*{Data Availability}
Data used for this study can be accessed via the following link: \url{https://zenodo.org/records/10578271}.

\section*{Acknowledgements}
Funding for this work is provided through the USC-ISI Exploratory Research Award, NSF (award \#2051101), and through DARPA (awards \#HR0011260595 and \#HR001121C0169).
\bibliography{citations}

\begin{thebibliography}{34}
\providecommand{\natexlab}[1]{#1}

\bibitem[{Anand(2022)}]{anand2022twitter}
Anand, J. 2022.
\newblock Twitter doubles down on spam bots following Elon Musk's tweet.
\newblock \emph{India Today}.

\bibitem[{Chandrasekharan et~al.(2022)Chandrasekharan, Jhaver, Bruckman, and
  Gilbert}]{chandrasekharan2020quarantined}
Chandrasekharan, E.; Jhaver, S.; Bruckman, A.; and Gilbert, E. 2022.
\newblock Quarantined! Examining the Effects of a Community-Wide Moderation
  Intervention on {Reddit}.
\newblock \emph{CSCW}, 29(4).

\bibitem[{Chandrasekharan et~al.(2017)Chandrasekharan, Pavalanathan,
  Srinivasan, Glynn, Eisenstein, and Gilbert}]{Chandrasekharan2017}
Chandrasekharan, E.; Pavalanathan, U.; Srinivasan, A.; Glynn, A.; Eisenstein,
  J.; and Gilbert, E. 2017.
\newblock You Can't Stay Here: The Efficacy of {Reddit}'s 2015 Ban Examined
  Through Hate Speech.
\newblock \emph{CSCW}, 1(CSCW).

\bibitem[{Cresci et~al.(2017)Cresci, Di~Pietro, Petrocchi, Spognardi, and
  Tesconi}]{cresci2017paradigm}
Cresci, S.; Di~Pietro, R.; Petrocchi, M.; Spognardi, A.; and Tesconi, M. 2017.
\newblock The paradigm-shift of social spambots: Evidence, theories, and tools
  for the arms race.
\newblock In \emph{WWW}, 963--972.

\bibitem[{Cresci et~al.(2019)Cresci, Lillo, Regoli, Tardelli, and
  Tesconi}]{cresci2019cashtag}
Cresci, S.; Lillo, F.; Regoli, D.; Tardelli, S.; and Tesconi, M. 2019.
\newblock Cashtag piggybacking: Uncovering spam and bot activity in stock
  microblogs on Twitter.
\newblock \emph{TWEB}, 13(2): 1--27.

\bibitem[{Duffy and LeBlanc(2022)}]{duffy2022elon}
Duffy, C.; and LeBlanc, P. 2022.
\newblock Elon Musk restores Donald Trump's Twitter account.
\newblock \emph{CNN}.

\bibitem[{Eisenstein, Ahmed, and Xing(2011)}]{eisenstein2011sparse}
Eisenstein, J.; Ahmed, A.; and Xing, E.~P. 2011.
\newblock Sparse additive generative models of text.
\newblock In \emph{ICML-11}, 1041--1048.

\bibitem[{Ferrara et~al.(2016)Ferrara, Varol, Davis, Menczer, and
  Flammini}]{ferrara2016rise}
Ferrara, E.; Varol, O.; Davis, C.; Menczer, F.; and Flammini, A. 2016.
\newblock The rise of social bots.
\newblock \emph{Commun. ACM}, 59(7): 96--104.

\bibitem[{Frenkel and Conger(2022)}]{frenkel2022hate}
Frenkel, S.; and Conger, K. 2022.
\newblock Hate Speech’s Rise on Twitter Is Unprecedented, Researchers Find.
\newblock \emph{NYT}.

\bibitem[{Geoffrey~Leech(2001)}]{Leech2001}
Geoffrey~Leech, A.~W., Paul~Rayson. 2001.
\newblock \emph{Word Frequencies in Written and Spoken English: based on the
  British National Corpus}.
\newblock London: Longman.
\newblock ISBN 0582-32007-0.

\bibitem[{{Google Jigsaw}(2017)}]{Jigsaw2017}
{Google Jigsaw}. 2017.
\newblock Perspective API.

\bibitem[{Hine et~al.(2017)Hine, Onaolapo, De~Cristofaro, Kourtellis,
  Leontiadis, Samaras, Stringhini, and Blackburn}]{hine2017kek}
Hine, G.~E.; Onaolapo, J.; De~Cristofaro, E.; Kourtellis, N.; Leontiadis, I.;
  Samaras, R.; Stringhini, G.; and Blackburn, J. 2017.
\newblock Kek, cucks, and god emperor trump: A measurement study of {4Chan}’s
  politically incorrect forum and its effects on the web.
\newblock In \emph{ICWSM}.

\bibitem[{Horta~Ribeiro et~al.(2021)Horta~Ribeiro, Jhaver, Zannettou,
  Blackburn, Stringhini, De~Cristofaro, and West}]{horta2021platform}
Horta~Ribeiro, M.; Jhaver, S.; Zannettou, S.; Blackburn, J.; Stringhini, G.;
  De~Cristofaro, E.; and West, R. 2021.
\newblock Do platform migrations compromise content moderation? evidence from
  r/the\_donald and r/incels.
\newblock \emph{CSCW}, 5(CSCW2): 1--24.

\bibitem[{Johnson et~al.(2019)Johnson, Leahy, Restrepo, Vel{\'a}squez, Zheng,
  Manrique, Devkota, and Wuchty}]{johnson2019hidden}
Johnson, N.~F.; Leahy, R.; Restrepo, N.~J.; Vel{\'a}squez, N.; Zheng, M.;
  Manrique, P.; Devkota, P.; and Wuchty, S. 2019.
\newblock Hidden resilience and adaptive dynamics of the global online hate
  ecology.
\newblock \emph{Nature}, 573(7773): 261--265.

\bibitem[{Lewis, Rowe, and Wiper(2019)}]{lewis2019online}
Lewis, R.; Rowe, M.; and Wiper, C. 2019.
\newblock Online/offline continuities: Exploring misogyny and hate in online
  abuse of feminists.
\newblock In \emph{Online othering}, 121--143. Springer.

\bibitem[{Mac, Isaac, and McCabe(2022)}]{mac2022resignations}
Mac, R.; Isaac, M.; and McCabe, D. 2022.
\newblock Resignations Roil Twitter as Elon Musk Tries Persuading Some Workers
  to Stay.
\newblock \emph{NYT}.

\bibitem[{Marlow, Miller, and Roberts(2021)}]{marlow2021bots}
Marlow, T.; Miller, S.; and Roberts, J.~T. 2021.
\newblock Bots and online climate discourses: Twitter discourse on President
  Trump’s announcement of US withdrawal from the Paris Agreement.
\newblock \emph{Climate Policy}, 21(6): 765--777.

\bibitem[{O'Brien and Ortutay(2022)}]{obrien2022musk}
O'Brien, M.; and Ortutay, B. 2022.
\newblock Musk’s Twitter disbands its Trust and Safety advisory group.
\newblock \emph{Associated Press}.

\bibitem[{O'Sullivan(2022)}]{osullivan2022twitter}
O'Sullivan, D. 2022.
\newblock Twitter is no longer enforcing its Covid misinformation policy.
\newblock \emph{CNN Business}.

\bibitem[{O'Sullivan and Duffy(2022)}]{osullivan2022elon}
O'Sullivan, D.; and Duffy, C. 2022.
\newblock Elon Musk has taken control of Twitter and fired its top executives.
\newblock \emph{CNN Business}.

\bibitem[{Saha, Chandrasekharan, and De~Choudhury(2019)}]{saha2019prevalence}
Saha, K.; Chandrasekharan, E.; and De~Choudhury, M. 2019.
\newblock Prevalence and psychological effects of hateful speech in online
  college communities.
\newblock In \emph{WebSci}, 255--264.

\bibitem[{Sato(2022)}]{sato2022buying}
Sato, M. 2022.
\newblock Buying Twitter ‘is not a way to make money,’ says Musk in TED
  interview.
\newblock \emph{The Verge}.

\bibitem[{Saveski, Roy, and Roy(2021)}]{saveski2021structure}
Saveski, M.; Roy, B.; and Roy, D. 2021.
\newblock The structure of toxic conversations on Twitter.
\newblock In \emph{WWW}, 1086--1097.

\bibitem[{Sayyadiharikandeh et~al.(2020)Sayyadiharikandeh, Varol, Yang,
  Flammini, and Menczer}]{sayyadiharikandeh2020detection}
Sayyadiharikandeh, M.; Varol, O.; Yang, K.-C.; Flammini, A.; and Menczer, F.
  2020.
\newblock Detection of novel social bots by ensembles of specialized
  classifiers.
\newblock In \emph{CIKM}, 2725--2732.

\bibitem[{Schmitz, Burghardt, and Muric(2022)}]{Schmitz2022}
Schmitz, M.; Burghardt, K.; and Muric, G. 2022.
\newblock Quantifying How Hateful Communities Radicalize Online Users.
\newblock In \emph{ASONAM}, 139--146.

\bibitem[{Sherman and Thomas(2022)}]{sherman2022elon}
Sherman, N.; and Thomas, D. 2022.
\newblock Elon Musk strikes deal to buy Twitter for \$44bn.
\newblock \emph{BBC}.

\bibitem[{Shi et~al.(2020)Shi, Liu, Yang, Zhang, Wen, and Su}]{shi2020social}
Shi, W.; Liu, D.; Yang, J.; Zhang, J.; Wen, S.; and Su, J. 2020.
\newblock Social bots’ sentiment engagement in health emergencies: A
  topic-based analysis of the COVID-19 pandemic discussions on Twitter.
\newblock \emph{IJERPH}, 17(22): 8701.

\bibitem[{Soucy(2022)}]{GlobalNews2022}
Soucy, P. 2022.
\newblock Trucker convoy leaves Kingston heading for weekend rally in Ottawa.
\newblock \emph{Global News}.

\bibitem[{Tiffany(2022)}]{tiffany2022are}
Tiffany, K. 2022.
\newblock Are the Libs Really Leaving?
\newblock \emph{The Atlantic}.

\bibitem[{Uyheng and Carley(2020)}]{uyheng2020bots}
Uyheng, J.; and Carley, K.~M. 2020.
\newblock Bots and online hate during the COVID-19 pandemic: case studies in
  the United States and the Philippines.
\newblock \emph{Journal of computational social science}, 3(2): 445--468.

\bibitem[{Varol et~al.(2017)Varol, Ferrara, Davis, Menczer, and
  Flammini}]{varol2017online}
Varol, O.; Ferrara, E.; Davis, C.; Menczer, F.; and Flammini, A. 2017.
\newblock Online human-bot interactions: Detection, estimation, and
  characterization.
\newblock In \emph{ICWSM}, volume~11, 280--289.

\bibitem[{Wang, Angarita, and Renna(2018)}]{wang2018era}
Wang, P.; Angarita, R.; and Renna, I. 2018.
\newblock Is this the era of misinformation yet: combining social bots and fake
  news to deceive the masses.
\newblock In \emph{WWW}, 1557--1561.

\bibitem[{Yang, Ferrara, and Menczer(2022)}]{yang2022botometer}
Yang, K.-C.; Ferrara, E.; and Menczer, F. 2022.
\newblock Botometer 101: Social bot practicum for computational social
  scientists.
\newblock \emph{arXiv preprint arXiv:2201.01608}.

\bibitem[{Zannettou et~al.(2018)Zannettou, Bradlyn, De~Cristofaro, Kwak,
  Sirivianos, Stringini, and Blackburn}]{zannettou2018gab}
Zannettou, S.; Bradlyn, B.; De~Cristofaro, E.; Kwak, H.; Sirivianos, M.;
  Stringini, G.; and Blackburn, J. 2018.
\newblock What is {Gab}: A bastion of free speech or an alt-right echo chamber.
\newblock In \emph{WWW}, 1007--1014.

\end{thebibliography}

\end{document}